\documentclass[aps,prb,twocolumn,groupedaddress]{revtex4-1}
\pdfoutput=1

\usepackage{braket}
\usepackage{amsmath}
\usepackage{amssymb}
\usepackage{amsfonts}
\usepackage{euscript}
\usepackage{bm}
\usepackage{booktabs}
\usepackage{multirow}
\usepackage{mhchem}
\usepackage{color}

\newcommand{\vv}[1]{{\boldsymbol{{#1}}}}
\newcommand{\mub}{\mu_\text{B}}
\newcommand{\muba}{\mu_\text{B}/\text{\AA}^2}
\newcommand{\mubaa}{\mu_\text{B}\text{\AA}}
\newcommand{\tenmuba}{ {\times 10^{-3} \muba} }
\newcommand{\tenmubaa}{ {\times 10^{-3} \mubaa} }

\begin{document}
\title{First-principles calculation of the bulk magnetoelectric monopolization: Berry phase and Wannier function approaches}
\date{\today}

\author{Florian Th\"ole}
\author{Michael Fechner}
\author{Nicola A. Spaldin}
\affiliation{Materials Theory, ETH Zurich, Wolfgang-Pauli-Strasse 27, 8093 Z\"urich, Switzerland}

\begin{abstract}
We present a formalism to calculate the macroscopic magnetoelectric monopolization from first principles within the density functional theory framework. An expression for the monopolization in the case of insulating collinear magnetism is derived first in terms of spin-polarized Wannier functions then recast as a Berry phase. We propose an extension to the general, non-collinear case which we implement computationally in the Wannier function form and use to calculate the magnetoelectric monopolizations of \ce{LiMnPO4} and \ce{Cr2O3}. We find that, while the former is well approximated by a summation over the formal local spin moments, the latter shows significant deviations from this approximation. We suggest that equating the Berry phase value with a sum over local moments provides an unambiguous route to defining the size of the local magnetic moment in magnetoelectric antiferromagnets containing only one type of magnetic ion. 
\end{abstract}

\maketitle

\section{Introduction}

The interaction energy, $H_{\rm int}$, of a magnetization density $\vv{\mu}(\vv{r})$ with an inhomogeneous magnetic field $\vv{H}\left(\vv{r}\right)$ can be written as a multipole expansion in powers of field gradients calculated at some arbitrary reference point $\vv{r}= 0$:
\begin{align} 
\label{eq:inhomog}
H_{\rm int} = &  -  \mu_0 \int \vv{\mu}(\vv{r}) \cdot \vv{H}\left(\vv{r}\right) d^3 \vv{r} \nonumber \\
            = &  - \mu_0 \int \vv{\mu}(\vv{r}) \cdot \vv{H}\left(0\right) d^3 \vv{r} \nonumber \\
 &-  \mu_0 \int r_{i} {\mu}_{j} (\vv{r}) \partial_{i} H_{j}\left(0\right) d^3 \vv{r} - \ldots.
\end{align}
where $i,j$ are Cartesian directions (summation over repeated indices is implied).\cite{Spaldin2008}

The first term, which is sufficient for describing many magnetic phenomena, gives the usual interaction of the magnetic dipole moment, $\vv{m} =  \int \vv{\mu} (\vv{r}) d^3 \vv{r}$, with a uniform magnetic field. 
Well-established methods exist for calculating the magnetic dipole moment and its energy within the density functional theory formalism:
For the case of the spin contribution to the magnetic moment, the relevant quantities are particularly straightforward to calculate, as the spin magnetic moment per unit volume (the magnetization) in collinear systems is simply the difference between the up- and down-spin charge densities which are directly accessible from a density functional calculation. \cite{Bousquet2011a}
The orbital magnetization is more complicated, since it is the expectation value of the circulation operator $\vv{r} \times \vv{v}$, which is not well defined in the Bloch representation. 
In spite of this difficulty, however, a formalism has also been recently developed for the calculation of orbital magnetization \cite{Thonhauser2005} and applied, for example, to the calculation of nuclear magnetic resonance (NMR) shielding tensors \cite{Thonhauser2009} and the orbital contribution to the magnetoelectric response \cite{Malashevich2010}.

In this work we present a formalism and initial results for the first-principles calculation of one component of the second terms, the so-called \emph{magnetoelectric multipoles}, in the multipole expansion.
Our motivation is many-fold. 
First, these terms are non-zero in materials that show a linear magnetoelectric response\cite{Ederer2007,Spaldin2008,Spaldin2013} and so can be used to classify and indeed to identify new magnetoelectric materials. 
Second, since they break both time-reversal and space-inversion symmetries, they offer candidate order parameters for completing the group of primary ferroics.
Currently ferromagnetism, ferroelectricity and ferroelasticity break time-reversal only, space-inversion only, and neither symmetries respectively, and a ferroic order which breaks both symmetries is sought.\cite{Spaldin2008, Ederer2007}
Indeed, the existence of ordered domains of such magnetoelectric multipoles was recently demonstrated using second-harmonic generation and hysteretic poling in \ce{LiCoPO4}.\cite{Zimmermann2014, VanAken2007}
In addition, since routine techniques for their direct measurement are not yet available, they represent a kind of ``hidden magnetic order'' beyond that of magnetic dipoles, analogous to the challenge presented by antiferromagnets a hundred years ago. 

Following earlier work,\cite{Spaldin2013} we decompose the nine-component tensor ${\cal M}_{ij} = \int r_{i} {\mu_j(\vv{r})}d^3 \vv{r}$ in Eq.~\ref{eq:inhomog} into three irreducible tensors, each of which changes sign under time-reversal and space-inversion symmetries. 
\begin{itemize}
\item[i)] the trace of the tensor, which couples to the divergence of the magnetic field, and so is often referred to as the monopole component:
\begin{equation}
a =  \frac{1}{3} {\cal M}_{ii} = \frac{1}{3} \int \vv{r} \! \cdot
\vv{\mu}(\vv{r}) d^3 \vv{r} \quad,
\label{Eqn_monopole}
\end{equation}
\\

\item[ii)] the toroidal moment vector dual to the antisymmetric part of the tensor:
$t_{i} = \frac{1}{2} \varepsilon_{ijk} {\cal M}_{jk}$,
which couples to the curl of the magnetic field,
\begin{equation} \label{eq:spinT}
\vv{t} = \frac{1}{2}  \int \vv{r} \! \times \vv{\mu}(\vv{r}) d^3 \vv{r} \quad,
\end{equation}
and
\\

\item[iii)] the traceless symmetric tensor $q_{ij}$ describing the magnetic quadrupole moment of
the system, which couples to the field gradient:
\begin{eqnarray}
q_{ij} &=& \frac{1}{2}\left({\cal M}_{ij} + {\cal M}_{ji} - \frac{2}{3} \delta_{ij} {\cal
M}_{kk}\right)\nonumber \\ &=& \frac{1}{2} \int \left[r_i \mu_j(\vv{r}) +
r_j \mu_i(\vv{r}) - \frac{2}{3} \delta_{ij} \vv{r}\! \cdot \vv{\mu(\vv{r})}
\right] d^3\vv{r} \quad.
\end{eqnarray}
\end{itemize}

The expansion of Eq.~(\ref{eq:inhomog}) can then be written in the form
\begin{eqnarray} \label{eq:inhomog3}
\mu_0^{-1}H_{\rm int} & = & - \vv{m} \cdot \vv{H}\left(0\right) \nonumber 
             - a \left(\nabla \cdot \vv{H}\right)_{\vv{r} = 0} \nonumber \\
            &   & - \vv{t} \cdot \left[ \nabla \times \vv{H} \right]_{\vv{r} = 0} \nonumber 
             - q_{ij} \left(\partial_{i} H_{j} + \partial_{j} H_{i}\right)_{\vv{r} = 0} - \ldots .
\end{eqnarray}
This decomposition transparently yields three terms that couple to the divergence, curl and gradient of the magnetic field, respectively.
We call the first the magnetoelectric monopole to avoid confusion with a true magnetic monopole, the second is referred to as the toroidal moment or anapole and the third is the magnetic quadruople.
By analogy with the bulk magnetization, their corresponding bulk quantites per unit volume are then the magnetoelectric monopolization, toroidization and quadrupolization. 
In this work we focus on the spin contribution to the magnetoelectric monopolization; for the case of the monopolization the orbital contribution is formally zero, since $\vv{\mu}_{\text{orb}}\propto \vv{r}\times\vv{v}$, and $\vv{r}\cdot(\vv{r}\times\vv{v}) = 0$. 
We also outline the developments required to calculate the toroidization and quadrupolization which will be the subject of future work.

The remainder of this paper is organized as follows: In Section \ref{sec:bp}, we derive an analytical formula for the spin contribution to the macroscopic magnetoelectric monopolization in insulating collinear antiferromagnets, both in terms of Wannier functions and expressed as a Berry phase, in a form that is already accessible in most existing density functional codes. 
We propose an extension of the formalism for the non-collinear case and show that, while not formally rigorous, it provides a practical route for extracting the monopolization in the case of antiferromagnets with spin canting.  
In Section~\ref{sec:comp-details}, computational details for our density-functional calculations are given.  
In Section~\ref{sec:results}, we compute the bulk magnetoelectric monopolization for two materials, \ce{LiMnPO4} and \ce{Cr2O3}, and compare our results to the previously used local-moment approximation, in which the integral in Eq.~\ref{Eqn_monopole} is replaced by a sum over local dipole moments at atomic sites.
In the final section, we argue that the magnetoelectric monopolization in magnetoelectric, antiferromagnetic insulators with only one magnetic type of atom can be used to define an effective magnetic moment. Finally, we discuss the connection between the magnetoeletric monopolization and the magnetoelectric response.

\section{Derivation of expression for the macroscopic magnetoelectric monopolization}
\label{sec:bp}

The macroscopic magnetoelectric monopolization, $A$, of a system of volume $V$, is given by
\begin{equation}
A = \frac{1}{3V} \int  \vv{r} \cdot \vv{\mu}(\vv{r}) \, d^3\vv{r}  \quad ,
\label{eq:a-integral}
\end{equation}
where the integral is over all space. \cite{Spaldin2013} 
In the case of a finite system, the integral can be performed directly and the magnetoelectric monopolization extracted without ambiguity. 
For the periodic, bulk solids that we consider here, however, the non-periodicity of the position operator poses problems analogous to those encountered in defining a ferroelectric polarization or an orbital magnetization in a bulk system.
We write the magnetization density in terms of the vector of Pauli matrices, $\vv{\sigma}$, and spinors, $\Phi_n(\vv{r})$, summed over the band index $n$:
\begin{equation}
\vv{\mu}(\vv{r}) = \mub \sum_n \Phi_n(\vv{r})^\dagger \vv{\sigma} \Phi_n(\vv{r}) \quad .
\label{eq:mu-orbitals}
\end{equation}
This gives the following expression for the magnetoelectric monopolization:
\begin{equation}
A = \frac{\mub}{3V} \sum_n \int \Phi_n(\vv{r})^\dagger \vv{\sigma} \cdot \vv{r} \, \Phi_n(\vv{r}) \, d^3\vv{r} \quad ,
\label{eq:a-integral-orbitals}
\end{equation}
which we use as the starting point for our implementation.

\subsection{Insulating collinear systems}
\label{sec:collinear}
For collinear spin systems, Eq.~\ref{eq:a-integral-orbitals} can be separated into two equations, one for each spin channel.
Choosing the quantization axis to be along $z$ gives 
\begin{equation}
  A = \frac{\mub}{3V}\left[ \sum_{n} \int \Phi_n^{\uparrow}(\vv{r})^\dagger z \Phi_n^{\uparrow}(\vv{r})  -  \sum_{n} \int \Phi_n^{\downarrow}(\vv{r})^\dagger z \Phi_n^{\downarrow}(\vv{r}) \right]
  \label{eq:collinear-orbitals}
\end{equation}
One can recognize each part as the definition of the ferroelectric polarization along the $z$ direction for the respective spin channel.
By analogy to Refs.~\citenum{Resta1992} and \citenum{King-Smith1993}, one can then write Eq.~\ref{eq:collinear-orbitals} for the case of a bulk periodic system as the Berry phase expression: 
\begin{align}
A =& \frac{\mub}{3V} \Biggl[ \sum_{n_{\uparrow}} \int d^3\vv{k} \braket{\mathcal{U}_{n\vv{k}}^{\uparrow} | \nabla_{k_z} | \mathcal{U}_{n\vv{k}}^{\uparrow} }  \nonumber \\
&-\sum_{n_{\downarrow}} \int d^3\vv{k} \braket{\mathcal{U}_{n\vv{k}}^{\downarrow} | \nabla_{k_z} | \mathcal{U}_{n\vv{k}}^{\downarrow} }
 \Biggr]  \quad ,
 \label{eq:bloch-updown}
\end{align}
where $\ket{\mathcal{U}^{\sigma}_{n\vv{k}}}$ is the cell-periodic part of the Bloch functions for spin channel $\sigma$.

Alternatively, one can rewrite the Berry phase expression for the magnetoelectric monopolization using Wannier functions, which can be chosen to be exponentially localized in the case of insulators,\cite{Brouder2007} and which we will see provide a particularly intuitive basis for an extension to non-collinear magnetic systems.
The transformation from Bloch functions $\ket{\psi_{n\vv{k}}}$ to Wannier functions is in general written as\cite{Marzari2012}
\begin{equation}
\ket{W_{n\vv{R}}}=\frac{V}{(2\pi)^3} \int_{\text{BZ}} d\vv{k} e^{-i\vv{k}\cdot\vv{R}} \sum_m U_{mn}^{\vv{k}} \ket{\psi_{m\vv{k}}} \quad,
\end{equation}
where $U_{mn}^{\vv{k}}$ is a unitary rotation matrix, and $\vv{R}$ is a lattice vector (in the following we take $\vv{R}=0$).
Also, in terms of the cell-periodic part of the Bloch functions $\ket{\mathcal{U}_{nk}}$, the expectation value of the position operator, usually termed the ``Wannier center'', is given by 
\begin{equation}
\braket{W_n | \vv{r}| W_n} = \int d^3\vv{k} \braket{\mathcal{U}_{n\vv{k}}  | \nabla_\vv{k} | \mathcal{U}_{n\vv{k}} } \quad .
\label{eq:wf-pos-op}
\end{equation}

In the case of collinear spin-polarized systems, the spin-up and spin-down manifolds can be treated separately and therefore, there is a separate set of Wannier centers for each spin channel. 
The expression for the magnetoelectric monopolization then reads
\begin{equation}
A = \frac{\mub}{3V} \sum_n \left[ \braket{W_n^{\uparrow}|r_\alpha|W_n^{\uparrow}} - \braket{W_n^{\downarrow}|r_\alpha|W_n^{\downarrow}} \right] \quad ,
\label{eq:collinear-wannier}
\end{equation}
where $\ket{W_n^{\uparrow}}$ and $\ket{W_n^{\downarrow}}$ are the Wannier functions for the up- (down-)spin channel.

Thus, $A$ for insulating collinear systems can be obtained using any standard first-principles code in which the Berry phase or Wannier function calculation of the polarization is implemented simply by taking the difference between the polarization for up- and down-spin bands.

One small conceptual complication arises when extracting a collinear monopolization from a standard code, due to the fact that $\vv{P}$ is a vector property, while $A$ is a scalar. 
In the former case,  the direction of the $k$-space derivative is a projection of the polarization onto the respective axis. 
In the latter case, however, the direction of the $k$-space derivative corresponds to the direction in which the Pauli matrix $\sigma$ is assumed to be diagonal, that is, the method assumes that the k-space derivative direction is the quantization axis of the collinear spin system, even if spin-orbit coupling is not included in the calculation. 
A standard density functional code will therefore automatically provide values for the monopolization for all three orientations of the collinear spin system in a single calculation. Those values corresponding to orientations other than the actual orientation of interest should then be disregarded.

\subsection{Extension to non-collinear systems}
\subsubsection{Formulation in terms of Wannier functions}

The Wannier function expression provides a conceptually appealing route to extending the formalism for the case of non-collinear spin systems.
In the case of periodic crystalline insulators, one can identify the spinors in Eq.~\ref{eq:a-integral-orbitals} with spinor Wannier functions $\ket{W_n}$.
 Then, after a switch to bra--ket notation, Eq.~\ref{eq:a-integral-orbitals} reads 
\begin{equation}
A = \frac{\mub}{3V} \sum_n \braket{W_n | \vv{\sigma} \cdot \vv{r} | W_n} \quad .
\label{eq:a-sum-wannier}
\end{equation}
While the ``ordinary'' Wannier center $\braket{W_n|\vv{r}|W_n}$ in the multi-band case is not gauge invariant, the sum over all Wannier centers is, so that the two terms in Eq.~\ref{eq:collinear-wannier} are rigorously well-defined. 
To provide a rigorous formal definition, the sum in Eq.~\ref{eq:a-sum-wannier} should also be invariant under gauge transformations among the Bloch states.
Our tests indicate that the gauge invariance of the sum of the Wannier centers transfers to the sum in Eq.~\ref{eq:a-sum-wannier}, although we do not have a formal proof of this.

In the case when all Wannier functions are internally collinear, the expression can be simplified to
\begin{equation}
A = \frac{\mub}{3V} \sum_n \braket{W_n | \vv{\sigma} | W_n} \cdot \braket{W_n | \vv{r} | W_n} \quad ,
\label{eq:a-sum-spin-r}
\end{equation}
where $\braket{W_n | \vv{\sigma} | W_n}$ is the expectation value of the spin of the Wannier function, giving both its magnitude and the orientation, and $\braket{W_n | \vv{r} | W_n}$ is the Wannier center. 
The expression above can then be directly compared to the local moment approximation employed in Ref.~\citenum{Spaldin2013}:
\begin{equation}
A = \frac{1}{3V} \sum_i \vv{m}_i \cdot \vv{R}_i  \quad ,
\label{eq:local-moment-approx}
\end{equation}
in which the local magnetic dipole moment of the $i$th ion, $\vv{m}_i$, replaces the spin of the Wannier function and the position of the $i$-th ion, $\vv{R}_i$, replaces the Wannier center.
It is clear from this comparison that in a fully ionic system, in which the magnetic moments are ``point spins'' located at the ionic sites, the Wannier centers will lie at the ionic sites and the two expressions will lead to identical values for the monopolization. 
In cases with covalency the two values will differ, just as the ferroelectric polarization in a covalent system differs from that in a point charge model.

We note that analogous Wannier function expressions for the toroidization $T_i$ and quadrupolization $Q_{ij}$ can be written as
\begin{equation}
{T}_i = \frac{\mub}{3V} \sum_n \epsilon_{ijk} \braket{W_n | \sigma_j  r_k | W_n}
\end{equation}
and
\begin{align}
{Q}_{ij} =& \frac{\mub}{3V} \sum_n \Bigl( 
\braket{W_n | \sigma_i  r_j | W_n}  
+\braket{W_n | \sigma_j  r_i | W_n} \nonumber \\  
 &- \frac{2}{3} \delta_{ij} \braket{W_n | \sigma_i  r_i | W_n} \Bigr) \quad, 
\end{align}
respectively.

\subsubsection{Formulation as a Berry phase}
By using the transformation from Wannier functions to the cell-periodic part of Bloch functions, Eq.~\ref{eq:a-sum-wannier} can be rewritten as
\begin{eqnarray}
A  & = & \frac{\mub}{3V} \sum_n \braket{W_n | \vv{\sigma} \cdot \vv{r}| W_n} \\
   & = & \frac{\mub}{3V} \sum_n \int d^3\vv{k} \braket{\mathcal{U}_{n\vv{k}} | \vv{\sigma} \cdot \nabla_\vv{k} | \mathcal{U}_{n\vv{k}}} \, .
\label{eq:wf-a-op}
\end{eqnarray}
Note that this expression suffers from the same ambiguity regarding the gauge dependence as we pointed out above for the
Wannier function case.
Interestingly, Batista et al.\cite{Batista2008} arrived earlier at a similar result which they interpreted 
only for the ferrotoroidic case, although in principle their derivation is also applicable to the monopolization.

\subsection{Multi-valuedness of the magnetoelectric monopolization}

It is clear from Eq.~\ref{eq:collinear-wannier} that the magnetoelectric monopolization is a multi-valued quantity, since the center of a Wannier function is only defined up to a lattice vector, $\vv{R}_\alpha$. 
The term ``monopolization increment'', $\Delta A$, was introduced in Ref.~\citenum{Spaldin2013} to describe the difference between branches of the associated monopolization lattice.
For the case of collinear spin systems, denoting the spin quantization axis of the system by $\vv{\sigma}$, the monopolization increment is
\begin{equation}
  \Delta A = \frac{\mub}{3V} \ \vv{\sigma} \cdot \vv{R}_\alpha \quad .
  \label{eq:quantum}
\end{equation}

For noncollinear systems the situation is more subtle, since in principle, each Wannier function in Eq.~\ref{eq:a-sum-wannier} can have a different spin direction. 
Therefore, the monopolization increment can be different for each Wannier function:
\begin{equation}
  \Delta A_n = \frac{\mub}{3V} \ \vv{\sigma}_n \cdot \vv{R}_\alpha \quad ,
  \label{eq:quantum-wannier}
\end{equation}
where $\vv{\sigma}_n$ is the vector describing the spin of the $n$-th Wannier function and multiple monopolization increments can exist.

\begin{figure}
\includegraphics{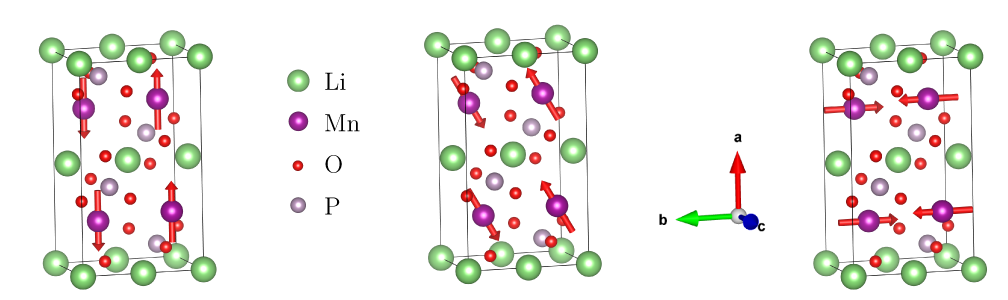}
\includegraphics{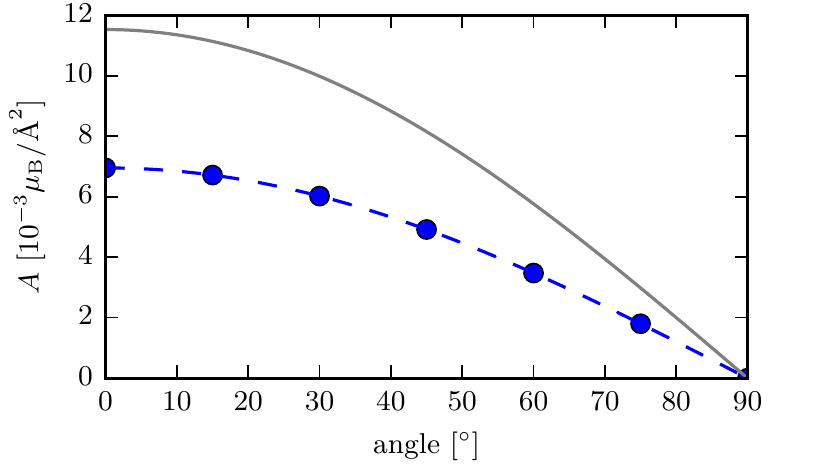}
\caption{Magnetoelectric monopolization in \ce{LiMnPO4} as a function of the angle of the spins to the $a$ axis. 
The monopolization has its maximum value when the spins are aligned along $a$, and drops to zero as they rotate away from the $a$ axis and their arrangement becomes toroidal. 
The gray line shows the monopolization increment. 
On top, the alignment of the spins along the crystal axis is shown for the direction which maximises the monopolization ($0^\circ$), an intermediate direction and a direction with zero monopolization ($90^\circ$).
}
\label{fig:angles}
\end{figure}

We see also that the monopolization increment changes as the direction of spins changes. 
This situation is illustrated in Fig.~\ref{fig:angles}, where we show the calculated evolution of the magnetoelectric monopolization and the monopolization increment as a function of the angular rotation of the spin direction away from that which maximises the magnetoelectric monopolization ($0^\circ$).
Noteworthy is the case when $\vv{\sigma}_n$ becomes almost perpendicular to $\vv{R}_\alpha$ and the monopolization tends to zero; then the increment also becomes small and changes between branches cannot be easily distinguished. 
Therefore, difficulties arise when one tries to define the monopolization difference between a non-monopolar reference structure and a monopolar structure if the change involves only the rotation of spins and no structural change. 

In this respect, the correspondence between the polarization and the monopolization is not exact. 
While both the electron charge and the electron spin have well-defined single values, the charge enters the expression for the polarization as a scalar quantity, whereas the spin enters as a vector, in a dot product with its position. 
This means that the polarization quantum is unchanged (provided that the lattice vectors are unchanged) even if the atomic positions evolve, whereas, as we have just seen, the monopolization increment evolves with the orientation of the spin moment. 

The situation is more straightforward in the case of a monopolization arising from a structural change. 
In this case, the monopolization increment is unchanged, in direct analogy to the polarization quantum.
An example of this case is \ce{FeS}, in which the transition from space group $P6_3/mmc$ to $P\bar{6}2c$ goes along with a loss of the inversion center and the occurence of a magnetoelectric monopolization.\cite{Ricci2015}

\section{Computational details}
\label{sec:comp-details}

Calculations presented here were performed using the Quantum Espresso code.\cite{Giannozzi2009}
We used the PBE functional and norm-conserving pseudopotentials, with $2s$ valence states for Li, $3d$ and $4s$  for Mn and Cr, $3s$ and $3p$ for P and $2p$ and $2s$ states for O.
For both \ce{Cr2O3} and \ce{LiMnPO4}, well-converged magnetoelectric monopolizations were obtained with an energy cutoff of~100 Ry for both total-energy and Berry phase calculations. 
The $k$-point grid was $3 \times 5 \times 5$ for \ce{LiMnPO4} and $4 \times 4 \times 2$ for \ce{Cr2O3} in hexagonal setting. 
For \ce{LiMnPO4}, we used the Hubbard $U$ correction\cite{Liechtenstein1995} on the Mn sites with $U=4$~eV and $J=0.5$~eV. 
Note that the allowed antiferromagnetic $A_z$-type canting of the Mn spins does not lead to an energy lowering in our calculations, so our system remains collinear. (Fortuitously, this allows for a direct comparison of our results with those of Ref.~\citenum{Spaldin2013}, where the canting was neglected.)

Collinear magnetoelectric monopolizations were obtained with the Berry phase implementation of Quantum Espresso.
To compare the Berry phase prescription and the Wannier function description of Eq.~\ref{eq:a-sum-spin-r}, we used the Wannier90 code\cite{Mostofi2014} to generate maximally localized Wannier functions. 

\section{Results}
\label{sec:results}

\subsection{\ce{LiMnPO4}}
\label{sec:res-bp}

We choose \ce{LiMnPO4}, which was shown previously to have a diagonal magnetoelectric response\cite{Mercier1969,Toft-Petersen2012}  and a corresponding macroscopic magnetoelectric monopolization,\cite{Spaldin2013} as our first model system.
The crystallographic space group of \ce{LiMnPO4} is \textit{Pnma} and
the antiferromagnetic $C_x$-type order of the Mn ions has the magnetic space group $Pn'm'a'$, which allows a macroscopic magnetoelectric monopolization.
In Ref.~\citenum{Spaldin2013}, the size of the monopolization obtained from summing over the localized magnetic moments as in Eq.~\ref{eq:local-moment-approx}, using the moments obtained from projecting into the muffin tin spheres (4.26 $\mub$), was found to be $5.89 \tenmuba$. In addition, a small contribution of $0.03 \tenmuba$ from summing the magnetoelectric monopoles in the spheres around each atom was found.
Using the lattice parameters and atomic positions from Ref.~\citenum{Spaldin2013}, we obtain a slightly smaller magnetic 
moment of $4.17~\mub$ on the Mn sites, which gives a correspondingly slightly smaller local-moment magnetoelectric monopolization of $A^{lm}=5.77 \tenmuba$. 
Note that all results from this section and the next are summarized in Tab.~\ref{tab:results}.

We begin by using the Berry phase formalism to calculate the macroscopic magnetoelectric monopolization of \ce{LiMnPO4}. 
As expected for an antiferromagnetic system in which the crystal structure contains inversion symmetry, we obtain values of equal magnitude, but opposite sign for the two spin channels along the three crystal axes. 
Their sum gives the polarisation, which is zero in all directions.
Their difference is different from the monopolization quantum in the $a$ direction, corresponding to the case where the Pauli matrix in this direction is diagonal, that is, the magnetic spins are aligned along $a$. 
The resulting monopolization is $A=6.945\tenmuba \pm n \Delta A$, where the monopolization increment $\Delta A=11.53\tenmuba$. For spin directions along $b$ and $c$, as found in the Fe, Co and Ni analogues of \ce{LiMnPO4}, the Berry phase magnetoelectric monopolization is equal to zero or the monopolization increment.

We now compare this result from the Berry phase calculation to the Wannier function formalism outlined above. 
Still in the collinear framework, we choose the Mn $d$ orbitals and O $p$ orbitals as projections and carry out the Wannier transformation in Wannier90. 
The numerical result for $A$ is almost unchanged ($A=6.953 \tenmuba$).
While this might seem trivial, it indicates that our choice of Wannier function projection has captured all relevant hybridizations that contribute to the monopolization. Including spin-orbit coupling changes the monopolization by only $10^{-6} \muba$. 

We see that in the case of LiMnPO$_4$ use of the local-moment approximation to calculate the magnetoelectric monopolization
makes an underestimate of $\sim$15 \% compared to the full calculation, when the dipole moment projected into the atomic 
sphere is taken as the local magnetic moment. Interestingly, when we take the full formal spin-only moment of $5~\mub$ for \ce{Mn^2+} 
we arrive at a local moment monopolization very close to the $A^{lm}=6.917 \tenmuba$, and
the difference between the local moment magnetoelectric monopolization and the full Berry phase 
magnetoelectric monopolization $\Delta A = 0.03 \tenmuba$ corresponds precisely to the contribution of the atomic site terms obtained previously.\cite{Spaldin2013} We suggest, however, that this intriguing correspondence is likely coincidental and wait to discuss it further
until after our analysis of Cr$_2$O$_3$ which follows.

\begin{table}
\caption{\label{tab:cr2o3}Lattice parameters and Wyckhoff posisitions for \ce{Cr2O3}. Values were obtained by structural relaxation using the PBE exchange-correlation functional. Experimental values are taken from Ref.~\citenum{Sawada1994}.}
\label{tab:results}
\begin{ruledtabular}
\begin{tabular}{llrrr}
& & & DFT & experiment \\
\hline
a [$\text{\AA}$] & &  & 4.962 		& 4.9570 \\
c [$\text{\AA}$] & &  & 13.570		& 13.5923 \\
Cr & 12c & $z$ & 0.348	& 0.348\\ 
O & 18e & $x$ & 0.304  & 0.306 \\ 
\end{tabular} 
\end{ruledtabular}
\end{table}

\begin{table}
\caption{Summary of the calculated magnetoelectric monopolizations for \ce{LiMnPO4} and \ce{Cr2O3}. The upper panel gives the formal and DFT local spin magnetic moment and corresponding local moment monopolizations $A^{\text{lm}}$, whereas the bottom values give the calculated Berry phase magnetoelectric monopolization and the derived effective moment.}
\begin{ruledtabular}
\begin{tabular}{lrr}
 & \ce{LiMnPO4} & \ce{Cr2O3} \\
\hline
$m^{\text{formal}} \,[\mub] $ 				&	5 & 3 		 		\\
$m^{\text{DFT}} \,[\mub] $ 					&	4.17 & 2.27		\\
$A^{\text{lm,formal}} \,[10^{-3} \muba]$ 	& 6.917 & 7.388		\\
$A^{\text{lm,DFT}}\,[10^{-3} \muba]$ 		&5.77 & 5.173		\\
\hline
$A^{\text{BP}}\,[10^{-3} \muba]$ 	& 6.945 & 5.488		\\
$m^{\text{eff}} \,[\mub] $ 			& 5.02			& 2.25 		\\
\end{tabular} 
\end{ruledtabular}
\end{table}

\subsection{\ce{Cr2O3}}

Next, we turn our attention to chromium dioxide, \ce{Cr2O3}, which is the prototypical magnetoelectric material.\cite{Dzyaloshinskii1959}
It crystallizes in the corundum structure (space group $R\bar{3}c$) with a collinear antiferromagnetic structure consisting of
antiferromagnetic chains along the hexagonal $c$ axis with the moments aligned along $c$. This leads to the magnetic space group 
$R\bar{3}'c'$ which allows a macroscopic monopolization but no spin canting. 
First, we fully relaxed the structure; the resulting coordinates are given in Table \ref{tab:cr2o3} alongside experimental values
with which they compare favorably.

The collinear Berry phase calculation along $z$ yields $A=5.488 \tenmuba$; using the Wannier function formalism with Cr $t_{2g}$ and O $p$ orbitals as projections we again obtain an almost unchanged result ($A=5.469 \tenmuba$).
Including spin-orbit coupling, one obtains $A=5.496 \tenmuba$.
While this change is not huge, the effect that spin-orbit coupling has on the monopolization is larger than in \ce{LiMnPO4}.

In our first-principles calculations, we obtain a magnetic moment of $2.27\,\mub$ on the Cr sites which gives a local 
moment magnetoelectric monopolization of $A=5.173\tenmuba$. As in the case of \ce{LiMnPO4} this is an underestimate of
the full Berry phase value, but this time by only $\sim$6\%. 
The correction from the atomic site monopoles is $0.69 \tenmuba$; adding this contribution to the local moment approximation leads to an overestimation of the Berry phase value by $\sim$7\%.
Taking the formal spin-only moment of $3\,\mub$ for \ce{Cr^3+}, however, one obtains $A~=~7.388\tenmuba$ which is a 
substantial overestimate of the Berry phase value. To recover the Berry phase result in a local-moment picture, 
one has to take an effective magnetic moment of $2.253\, \mub$ on the Cr sites.

\section{Discussion}

The origin of the differences between \ce{LiMnPO4} and \ce{Cr2O3}, in terms of the size of the local magnetic moment that
must be used to bring the monopolization calculated within the local moment approximation into agreement with the full 
Berry phase value, is unclear. In both cases, the local magnetic moments differ substantially from the formal ionic 
values (by 17\% in the case of \ce{LiMnPO4} and by 24\% for \ce{Cr2O3}), indicating significant hybridization between
oxygen $2p$ and transition metal $3d$ electrons, and consequent deviation from the ionic limit. It is therefore
particularly surprising that in \ce{LiMnPO4}, the local moment approximation is almost exact for the formal spin-only 
Mn moment of $5 \mub$. 
In both cases the contribution from the local monopolizations within the spheres around the atomic sites is small compared to the local moment monopolization.
In \ce{LiMnPO4}, it is negligible ($A^{\text{as}} = 0.03 \tenmuba$), owing largely to a cancellation of the contributions from different oxygen sites.\cite{Spaldin2013}
It is larger in \ce{Cr2O3} ($A^{\text{as}} = 0.69 \tenmuba$),
since while
the magnitudes of the atomic site oxygen contribution ($9.3\tenmubaa$) and Cr contribution ($2.6\tenmubaa$) are comparable to the analogous values in \ce{LiMnPO4}, all contributions have the same sign and therefore do not cancel each other.

Definition of the local magnetic moment associated with an ion is of course ambiguous in a covalently bonded solid,
and the differences likely reflect as much the details of the projection of the Bloch states into the atomic sphere
as real physics. We suggest that the value of local magnetic moment that brings the local-moment approximation to the 
magnetoelectric monopolization into equality with the full Berry phase value can be used as a way of unambiguously
defining the local magnetic moment in magnetoelectric antiferromagnets containing only one kind of magnetic ion, and 
it is certainly a relevant 
definition in the discussion of magnetoelectric monopolizations. Note that the local magnetic moment defined
in this way is distinct from the magnetic charge of Ref.~\onlinecite{Ye2014}, which gives
the change in magnetization with atomic displacement.

Next, we turn our attention to the relation between the magnetoelectric coefficients and the magnetoelectric monopolization. In Ref.~\citenum{Spaldin2013}, a relation between these two quantities was derived from the following free energy expression:
\begin{align}
  U = &\frac{1}{2\epsilon}P^2 - \vv{P}\cdot\vv{E} + \frac{\mu_0}{2\chi} M^2 - \mu_0 \vv{M} \cdot \vv{H} \nonumber \\
  &+ \frac{1}{2}\beta A^2 + \frac{1}{4}\gamma A^4 + c A \vv{P} \cdot \vv{M} \quad ,
\end{align}
where $\epsilon$ and $\chi$ are the electric and magnetic susceptibilities, $\beta$ and $\gamma$ are temperature-dependent coupling coefficients and $c$ determines the strength of the magnetoelectric coupling. 
Minimizing this expression leads to a relationship between the magnetoelectric coefficient and the monopolization with the 
electric and magnetic susceptibilities as proportionality constants:
\begin{equation}
\alpha = c \epsilon \chi A \quad.
\label{eq:alpha-c}
\end{equation}
The coupling strength $c$, however, is unknown. Indeed it was argued in Ref.~\onlinecite{Spaldin2008} that $c$ is not
of physical relevance, since the magnetoelectric tensor describes the second-order correction to the free energy 
in external electric and magnetic fields, while magnetic multipoles are generated by the expansion of the first-order 
correction to the free energy in powers of magnetic field gradients. 

Here, we estimate the value of $c$ for \ce{Cr2O3} using literature values for the various quantities appearing in Eq.~\ref{eq:alpha-c} and then make a comparison to calculated values in the high- and low-frequency limits. 
The results are summarized in Tab.~\ref{tab:coupling-constants}
The dielectric susceptibility is around 10 at low temperature, with only a small anisotropy between parallel and perpendicular (to $z$)
orientation.\cite{Fang1963, Lucovsky1977} 
The low temperature magnetic susceptibility in perpendicular orientation has the value $1.5 \times 10^{-3}$ (dimensionless SI).\cite{Foner1963}
From Ref.~\citenum{Hehl2008}, we take the monopolar part of the magnetoelectric tensor as $\widetilde{\alpha} = \frac{1}{3} \text{Tr}(\alpha) = 0.7 \text{ ps/m}$.
Putting together these experimental results and our calculated magnetoelectric monopolization, we estimate the coupling strength as $c = 1 \times 10^{-5} \text{ s/(Am)}$.

\begin{table}
\caption{Summary of the calculated coupling strengths $c$. Here, $\widetilde{\alpha}=\frac{1}{3}\,\text{Tr}(\alpha)$, and $\epsilon$  and $\chi$ are the electronic and magnetic susceptilibty.  Experimental values are taken from Refs.~\citenum{Hehl2008, Fang1963, Lucovsky1977, Foner1963} and the DFT results for $\widetilde{\alpha}$ and $\chi$ are taken from Ref.~\citenum{Bousquet2011a}.}
\label{tab:coupling-constants}
\begin{ruledtabular}
\begin{tabular}{llrrrr}
& & $\widetilde{\alpha}$ [ps/m] & $\epsilon$ & $\chi$ & $c$ [s/(Am)]  \\
\hline
\multicolumn{2}{c}{experimental } & 		0.7 	& 10 		& $1.5 \times 10^{-3}$ 	& $1\times 10^{-5}$ 	\\
\hline
\multirow{2}{*}{DFT} & electronic & 		0.23 	& 5.8 	& $1.9 \times 10^{-3}$ 		& $4\times 10^{-6}$ 	\\
& total & 										0.97 	& 9.3 	& $1.9 \times 10^{-3}$ 		& $1\times 10^{-5}$ \\
\end{tabular} 
\end{ruledtabular}
\end{table}

First-principles-based investigations are able to distinguish between electronic and lattice contributions to the various susceptibilites.
In the following we take the results from Ref.~\citenum{Bousquet2011a}
(note that a range of slightly different values for the spin-only response have been obtained,\cite{Malashevich2012, Bousquet2011a, Iniguez2008} with all values overestimating the experimental values)
 and augment them with new calculations of the static and high-frequency dielectric susceptibility. 
 By employing density functional pertubation theory, we obtain the lattice contribution $\epsilon^{\text{latt}} = 3.5$ and the electronic contribution $\epsilon^{\text{el}} = 5.8$. 
 The two contributions agree reasonably well with the experimental values.\cite{Lucovsky1977} 
 The lattice contribution to the spin magnetic susceptibility has been previously shown to be negligible.\cite{Ye2014}

Since $A$ is a thermodynamic quantity, that is, it is not frequency dependent and $\chi$ is almost frequency independent, we see immediately that the proportionality factor 
\begin{equation}
  c = \frac{\alpha}{\epsilon \chi A}
\end{equation}
 is only constant if the frequency dependence of $\epsilon$ and $\alpha$ is the same.
The spin-electronic magnetoelectric response leads to a high-frequency (electronic only) coupling strength of $c^{\infty}=4 \times 10^{-6} \text{ s/(Am)}$, while the total spin magnetoelectric response leads to a low-frequency (electronic plus lattice) coupling strength $c^{\text{tot}}=1\times 10^{-5} \text{ s/(Am)}$. 
The latter value is consistent with the estimates extracted from the experimental range of susceptibilites and the difference between $c^{\text{tot}}$ and $c^{\infty}$ confirms the assertion of Ref.~\citenum{Spaldin2008} that the proportionality constant does not represent a fundamental, physically universal parameter.
Indeed, we expect the behaviour of $c$ to be especially interesting close to magnetoelectric phase transitions, where it is known that $\epsilon$, $\chi$ and $\alpha$ can all diverge.\cite{Dzyaloshinskii2011, Bousquet2011}

\section{Summary}
In summary, we have derived the Berry phase theory for the macroscopic magnetoelectric monopolization for insulating collinear antiferromagnets, proposed a generalization to the non-collinear phase, and implemented it in its Wannier function form within the density functional formalism. 
We applied the method to two prototypical magnetoelectric materials, \ce{LiMnPO4} and \ce{Cr2O3}.
Our results highlight two different behaviours: In \ce{LiMnPO4}, the bulk monopolization is close to the value obtained by a simple local moment formalism using the formal ionic magnetic moment. 
In contrast, in \ce{Cr2O3}, use of the projected atomic-site local moment yields a better agreement. 
We proposed comparison of the local-moment approximation and full Berry phase values of the monopolization as an unambiguous way to define the local magnetic moment in magnetoelectric antiferromagnets. 
Finally we discussed the quantitative connection between the magnetoelectric response and the monopolization via the dielectric and magnetic susceptibilities. 

\section{Acknowledgments}
This work was supported financially by the ETH Z\"urich (NAS), by the ERC Advanced Grant program, No.~291151 (MF and NAS), by the Max R\"ossler Price of the ETH Z\"urich (NAS), and by the Sinergia program of the Swiss National Science Foundation Grant No. \mbox{CRSII2\_147606/1} (FT and NAS). This work was supported by a grant from the Swiss National Supercomputing Center (CSCS) under project IDs s624 and p504. We thank Eric Bousquet, Lars Nordstr\"om and David Vanderbilt for fruitful discussions.

\bibliography{lib_clean_2016-02-16}

\end{document}